\documentclass{aastex631}
\renewcommand{\baselinestretch}{1.25}
\usepackage[utf8]{inputenc}
\usepackage{mathtools}
\usepackage{nameref}
\usepackage{hyperref}
\usepackage{booktabs}
\usepackage{float}
\usepackage{enumitem}
\usepackage{xcolor}

\begin{document}

\title{The fuzzy cores of Jupiter and Saturn}
\author[0000-0001-5555-2652]{Ravit Helled}
\affiliation{Department of Astrophysics, University of Zürich, \\
             Winterthurerstrasse 190, 8057 Zürich, Switzerland}
\email{ravit.helled@uzh.ch}
\author[0000-0001-9432-7159]{David J. Stevenson}
\affiliation{Division of Geological and Planetary Sciences, California Institute of Technology, \\
             Pasadena, California 91125, USA}
\correspondingauthor{Ravit Helled}

\begin{abstract}
New interior models of Jupiter and Saturn suggest that both planets have “fuzzy cores”. These cores should be viewed as central regions that are enriched with heavy elements but are not distinct from the rest of the deep interior. These cores may contain large amounts of hydrogen and helium though small pure-heavy element cores may also exist.  New measurements along with advanced planetary modeling have revolutionized the way we think about the interiors of giant planets and provide important constraints for planet formation and evolution theories. These developments are also relevant for the characterization of giant exoplanets.
\end{abstract}

\section*{Plain Language Summary}
For decades giant planets were assumed to have distinct heavy-element cores in their deep interiors overlain by an envelope of hydrogen and helium. 
Measurements from the Juno and the Cassini missions strongly indicate that the internal structure of Jupiter and Saturn is much more complex, including inhomogeneities in composition and``fuzzy" cores. In Jupiter, this dilution might extend to more than one hundred Earth masses, while in Saturn it is  less, though both planets have a diluted region that extends to many tens of percent of total radius. In both cases, a small central core of pure or nearly pure elements may exist. These structures pose challenges for our understanding of giant planet formation and evolution.

\section*{Key Points}
\begin{itemize}
\item  The interiors of Jupiter and Saturn are complex; they  include  composition gradients and``fuzzy" cores.
\item The origin of the ``fuzzy cores" is unknown, but is probably linked to the planetary formation process.  
\item Many open questions remain. Future modeling, observations, experiments, and links exoplanets can improve our understanding of the gas giants. 
\end{itemize}
%
%
%
\newpage
 \section{Introduction}

Jupiter and Saturn are the largest planets in our solar system. These planets have been studied for many decades theoretically and observationally. 
It is important to understand the nature of these two gas giant planets since they provide key information on the  origin of our planetary system, the behavior of materials at high pressure and temperature, and on the diversity of giant planets \cite{Guillot2004,2020NatRP...2..562H,Helled2022,Helled2022b,Stevenson22, MiguelVazan2023,2023ASPC..534..947G}. \\ 
Historically, Jupiter and Saturn have been modeled assuming simplified internal structures where the existence of a massive heavy-element core in their centers was assumed based on existing planet formation theories. 
In such traditional structure models of Jupiter and Saturn, the internal temperature was estimated by assuming isentropic structures, or, more precisely, two to three isentropic layers, each of a different composition with negligible temperature jumps at interfaces. 
This view of an isentropic (adiabatic) interior would be valid if the planets consisted of a homogeneous "envelope" (everything except a core) surrounding a small core. \\

Naively one would think that constraining the core mass of the giant planet is important for discriminating between the core accretion model and the disk instability model for giant planet formation \cite{2017SSRv..213....5B}. However,  we now know that both formation models can lead to a large range of possible core masses and planetary compositions \cite{2014prpl.conf..643H,2021exbi.book...12H}.
Nevertheless, the fact that the average heavy-element (here we use the astrophysical convention: heavy elements are everything except hydrogen and helium) enrichment of both Jupiter and Saturn is higher relative to solar abundances favors the core accretion model.   
One can conclude that both planets are enriched with heavy elements because both  Jupiter and especially Saturn have smaller radii than one expects for bodies with a solar composition. This ``observation", which is actually a theoretical inference based on our understanding of hydrogen and helium at high pressures and limitations on the thermal effect for a mostly homogeneous body, does not require the existence of a central concentration of heavy elements, i.e., a core,  but has historically been interpreted that way, in part because of computational convenience or simplicity, but also in part because of the basic principle of the core accretion model in which a core first forms (somewhat analogous to the formation of the entire Earth) and is then enveloped by inflow of hydrogen-helium (H-He) gas that is accreted from the disk \cite{1980PThPh..64..544M,1996Icar..124...62P}. \\
 The old view of giant planets having pure heavy-element cores is no longer valid thanks to accurate gravity data from the Juno and Cassini missions and ring seismology
for Saturn. It is now clear that the internal structures of Jupiter and Saturn are complex, and that the planets are unlikely to be fully adiabatic (convective).  
However, studying the interior structure of giant planets, including the nature of their cores, is challenging because we cannot directly observe their interiors. Instead, we must rely on interior structure models and formation/evolution simulations that reconstruct the available measurements of the planetary  physical properties, such as their mass, radius, luminosities, and gravitational and magnetic fields.  Another important lesson recently learned is that Jupiter and Saturn are different, and that Saturn is not simply a smaller version of Jupiter. Below, we summarize the current understanding of the internal structure of the planets, focusing on their fuzzy cores, and discuss the key differences between Jupiter and Saturn, the remaining challenges in interior modeling, and the connection to giant exoplanets. 

\section{Updated interior models}

\subsection{Jupiter}
Our comprehension of Jupiter's interior underwent a significant reevaluation following precise measurements of its gravitational field by the Juno spacecraft \cite{2017Sci...356..821B}.  
 Accurate gravity data constrain the gravitational  gravitational harmonics which depend on the internal density distribution in the planet.  The lower order gravitational  even harmonics (e.g., $J_2$, $J_4$) are sensitive down to about 60\% of the planetary radius in Jupiter in Saturn,  while the higher order even  harmonics probe the density distribution in the planetary atmosphere. Odd harmonics correspond to  north-south asymmetries and are therefore linked to atmosphere dynamics. 
Accurate measurements of the gravitational harmonics combined with interior models that use equations of states of materials expected to be in the planet and a temperature profile (that depends on the heat transport mechanism), can be used to constrain the planetary internal structure. 
Remarkably, the enhanced quality of gravity data from the Juno mission raised numerous new questions about the internal structure of Jupiter, challenging conventional structure models.
The unprecedented  accuracy of the Juno gravity data prompted the development of more sophisticated structure models of Jupiter, incorporating inhomogeneous composition, composition gradients, and diluted (fuzzy) cores.  
Updated interior models of Jupiter that fit Juno data include a non-uniform distribution of heavy elements within Jupiter's interior, consisting of a deep metallic envelope that is enriched in heavy elements compared to its upper molecular envelope  \cite{wahl2017,vazan2018,Debras2019,nettelmann2021,miguel2022,militzer2022,howard2023_interior}. 
These new Jupiter structure models also indicate a potentially significant deviation of the temperature profile in the planetary deep interior from the adiabatic one \cite{vazan2018,Debras2019}. 
Furthermore, in these updated models, Jupiter's core is no longer considered to be a purely heavy-element central region with a density discontinuity at the core-envelope boundary. 
This fuzzy/diluted core could extend to a few tens of percents of Jupiter's total radius. Except for the possible presence of a heavy element-dominated innermost region (a ``compact core" of a few Earth masses at most), the composition of the fuzzy core might even be thought of as ``dirty hydrogen". 
Overall, the total heavy-element mass in Jupiter remains uncertain (and is model dependent). {The heavy-element mass inferred from adiabatic models is of the order of 20 Earth masses (M$_{\oplus}$)} while non-adiabatic models predict higher masses that can reach up to  60 M$_{\oplus}$ of heavy elements  \cite{wahl2017,vazan2018,Debras2019,nettelmann2021,miguel2022,militzer2022,howard2023_interior,2022arXiv220210046H}. 

\subsection{Saturn}
It has been known for decades that Saturn's observed luminosity exceeds predictions from fully convective models, indicating that the influence of helium rain and/or non-convective regions is significant in this planet. 
 The presence of composition gradients and/or boundary layers  affect the heat transport mechanism, which in turn influences   the planetary temperature and luminosity. 
The luminosity of a mostly homogeneous planet is readily understood as the decline of thermal energy (most of which happens in early history) but is complicated if there are composition gradients or helium rain because of the competing effects of gravitational energy change and possible storage or change in internal thermal energy that is not proportional to the planetary effective temperature.  
 The indications that Saturn's interior is non-adiabatic, as its observed luminosity is higher than predicted by adiabatic/homogeneous models provide important constraints for modeling Saturn's internal structure today. 
It is still unknown whether  Saturn's observed luminosity is a result of helium rain \cite{Mankovich2020}  and/or composition gradients 
\cite{Leconte2013}. 
Gravity field and ring seismology measurements from the Cassini spacecraft provide new constraints on Saturn's internal structure 
\cite{2019Sci...364.2965I,2019ApJ...879...78M,2021NatAs...5.1103M}. In the case of Saturn, ring seismology revealed oscillation modes that can only be explained by a stably-stratified region, given clear evidence that Saturn's interior is non-adiabatic (and not fully mixed). Further details on ring seismology can be found in \cite{Mankovich2020AGU} and references therein.

\cite{2019ApJ...879...78M} presented interior models of Saturn assuming distinct layers and considering different rotation periods and core sizes. This structure model includes a fuzzy core and helium rain, which results in a large region in Saturn's deep interior with composition gradients that is stable against convection. 
The core mass was inferred to be between 15 and 18 M$_\oplus$ with less than 5 M$_\oplus$ of heavy elements in the envelope. 
 \cite{2021NatAs...5.1103M} presented interior models  of Saturn that are consistent with both gravity and seismic data.  These models suggest that Saturn has a stable region against convection which extends to $\sim$60\% of Saturn's radius. Also, the interior is found to include  composition gradients and a moderately low central density of $\sim$6 g cm$^{-3}$, indicating the existence of a fuzzy core, a result that is consistent with empirical structure models of Saturn \cite{2020ApJ...891..109M}. The total heavy-element mass in Saturn was inferred to be $\sim$ 17 M$_\oplus$. 
A similar result was inferred from a 4-layer model with an estimated  total heavy-element mass of 12–18 M$_\oplus$, where the interior includes a small compact core surrounded by a fuzzy core \cite{2020A&A...639A..10N}.
\cite{nettelmann2021} explored Saturn's interior using different H-He equations of state (EoSs), finding that models with the EoS presented by \cite{Chabrier2019} (known as the CMS  EoS) result in an enriched envelope extending to up to 0.4 of Saturn's radius and compact heavy-element cores. A solution of a fuzzy core extending to 40\% of Saturn's radius was found when EoS perturbations were considered. 
Sketches of the internal structures of Jupiter and Saturn are presented in Figure 1. 


\subsection{Compact cores?}
Both Jupiter and Saturn could possess small, compact heavy-element cores in their centers. Unfortunately, constraining the existence and/or properties of such small cores is very challenging.  
 It is extremely hard to determine whether a compact core exists because its expected mass is small, of the order of a few M$_{\oplus}$. This low mass is not just small in comparison to the total planetary mass, but is also located in a region which is "unseen" by the gravity data.  It also contributes negligibly to the moment of inertia, this is because a five M$_{\oplus}$ core of about an Earth radius contributes of order one part in $10^4$ of the moment of inertia. The gravity harmonics impose very tight constraints on the acceptable values of the moment of inertia, the value of which can be determined geodetically through the observed precession of the rotation axis. Juno is measuring this very accurately for the first time, but not with sufficient accuracy to improve interior models.\\
While the expected mass of such pure heavy-element cores is tiny compared to the total planetary mass, it is important for constraining planet formation and evolution theory. Recent giant planet formation models predict that the deep interior of giant planets should include such compact cores \cite{Lozovsky2017,Helled2017,Helled2022}. These pure heavy-element cores are formed by accretion of solids (pebbles, planetesimals), and as long as the core mass is below a couple of M$_{\oplus}$, the small amount of H-He gas that is co-accreted is insufficient to prevent planetesimals or most of the pebble mass penetrating to the center (core).  
Indeed, several structure models of both Jupiter and Saturn include such compact cores \cite{howard2023_interior,miguel2022} but solutions without compact cores have also been  presented.  
A complexity arises from the fact that such primordial compact cores could be destroyed by strong convective mixing \cite{Mueller2020c} or by extreme events such as a giant impact \cite{Liu2019}, as we discuss below.


 \begin{figure}
  \begin{center}
    \includegraphics[scale=0.5]{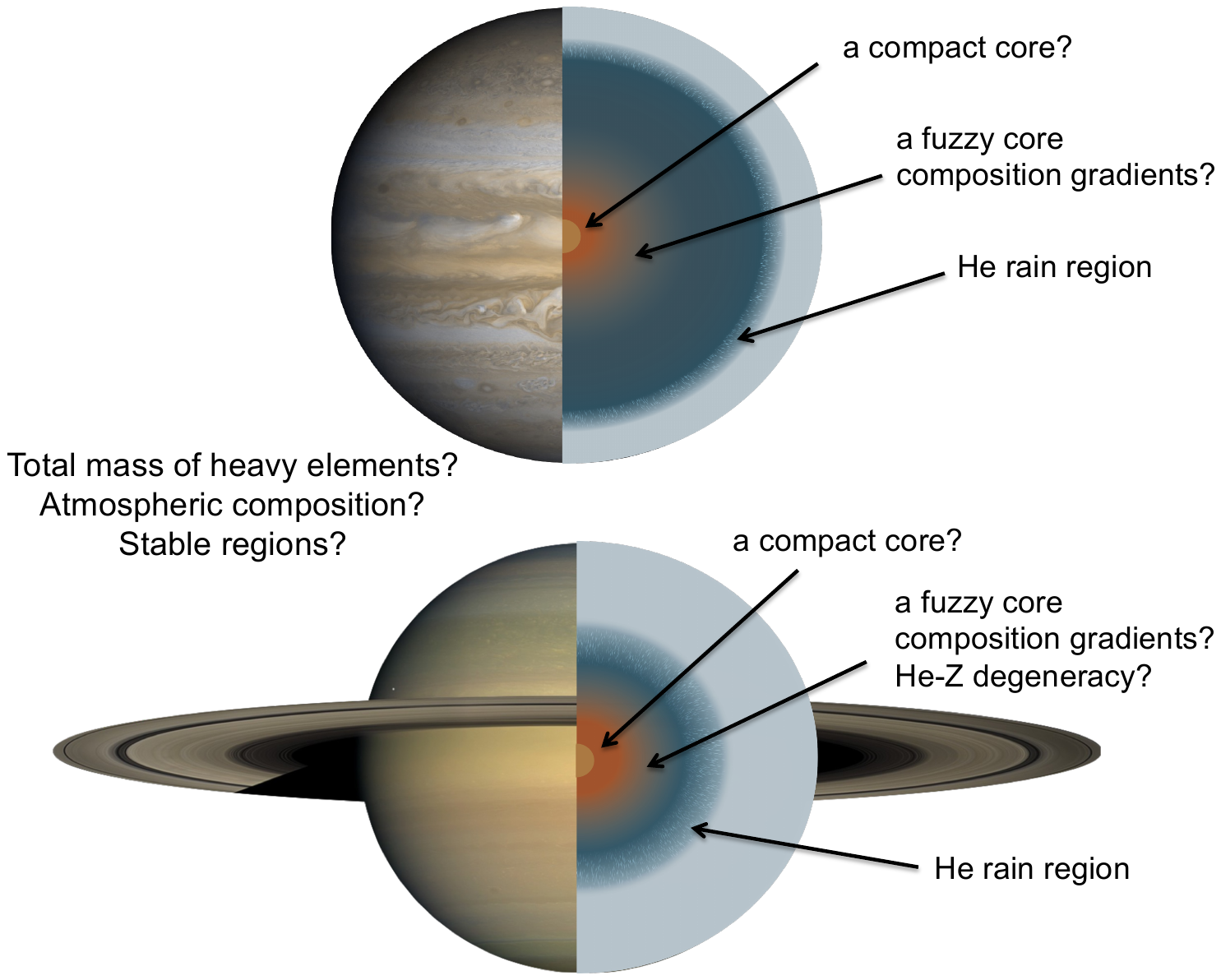}
    \caption{Sketches of the internal structures of Jupiter and Saturn. The two planets consist of a molecular hydrogen envelope where helium is depleted and a deeper region of helium rain where hydrogen becomes metallic at a pressure of about Mbar pressures. This region also includes heavy element but their concentration remains unknown. In the metallic region helium is enriched. The deeper interiors are expected to have fuzzy cores and/or composition gradients. The planetary deep interior has a larger concentration of heavy elements compared to the molecular envelope. Note that the layers are not necessarily distinct and the change between layers is gradual.  In the case of Saturn, it is hard to separate the contribution of helium and heavy elements in the deep interior (see text).  Both planets may still possess  small pure-heavy-element cores (section 2.3).}
    \label{fig:sketches}
  \end{center}
\end{figure}

\subsection{The difference between Jupiter and Saturn}
Although both Jupiter and Saturn are categorized as gas giant planets,  the two planets present clear and significant differences and Saturn should not be treated as being a smaller Jupiter. \\
First, in Saturn, helium demixing is presumably much more profound than in Jupiter due to the  mildly lower entropy for the outer region (including the observed atmosphere \cite{2020PSJ.....1...30A}), possibly leading to an inner region with less than 5\% hydrogen (e.g., \cite{2021NatAs...5.1103M}). The fact that much helium could exist in the deep interior of Saturn makes it harder to constrain the  heavy elements in the deep interior since a given density could be  explained by a larger amount of heavy elements or more helium. As a result, in the case of Saturn we have a heavy-element/helium degeneracy, which is difficult to resolve. However, a future measurement of Saturn's helium atmospheric abundance can constrain the process of helium rain in Saturn, and therefore could break this degeneracy \cite{2023arXiv230409215F}.  The H-He phase diagram also remains uncertain. Although this could be established by laboratory experiments, it is very difficult to determine the critical (unmixing) line of a mixture at Megabar pressures and many thousands of degrees K. \\
Second, the masses of Jupiter and Saturn are rather different with Saturn being more than a factor of three less massive. 
 The total heavy element masses in Jupiter and Saturn are comparable, perhaps a few tens of M$_{\oplus}$, implying that the metallicity of Saturn is higher. This is also the main reason why Saturn has a smaller radius than Jupiter. These two planets would have very similar radii if they had the same (but low) metallicity because of the approximate behavior of the H-He equation of state, wherein pressure scales roughly as density squared. \\
Third, the distribution of the heavy-elements within the planetary interiors is fundamentally different. This is demonstrated in Figure 2 which shows the heavy-element mass fraction (Z) in Jupiter and Saturn from various published models as described in the figure caption. Although the range of possible heavy-element distributions is large, two key conclusions can be made.  (1) The heavy-element mass fraction in the diluted cores of Jupiter and Saturn are significantly different. While for Jupiter the heavy-element mass fraction is of the order of 20\%, the fractions are significantly higher for Saturn. This might not be surprising given that the mass-radius relations of Jupiter and Saturn indicate that Saturn is more enriched in heavy elements than Jupiter in relative terms, but is important for noting that the nature of the fuzzy cores of Jupiter and Saturn are fundamentally different : Jupiter's diluted core is H-He-dominated in composition, while that of Saturn is not. (2) The mass included in the diluted core region is significantly different. Although model dependent, the mass of the dilute core is of the order of $\sim$100 M$_{\oplus}$ for Jupiter (about one third of the total planetary mass) comparable to the mass of Saturn; it is up to $\sim$60 M$_{\oplus}$ for Saturn.  
The different nature of the diluted cores of the planets may be linked to their formation history \cite{Helled2017,Helled2023}. 
We suggest that further efforts should be made to determine under what conditions such compact cores are required.

\begin{figure}
  \begin{center}
    \includegraphics[scale=0.42]{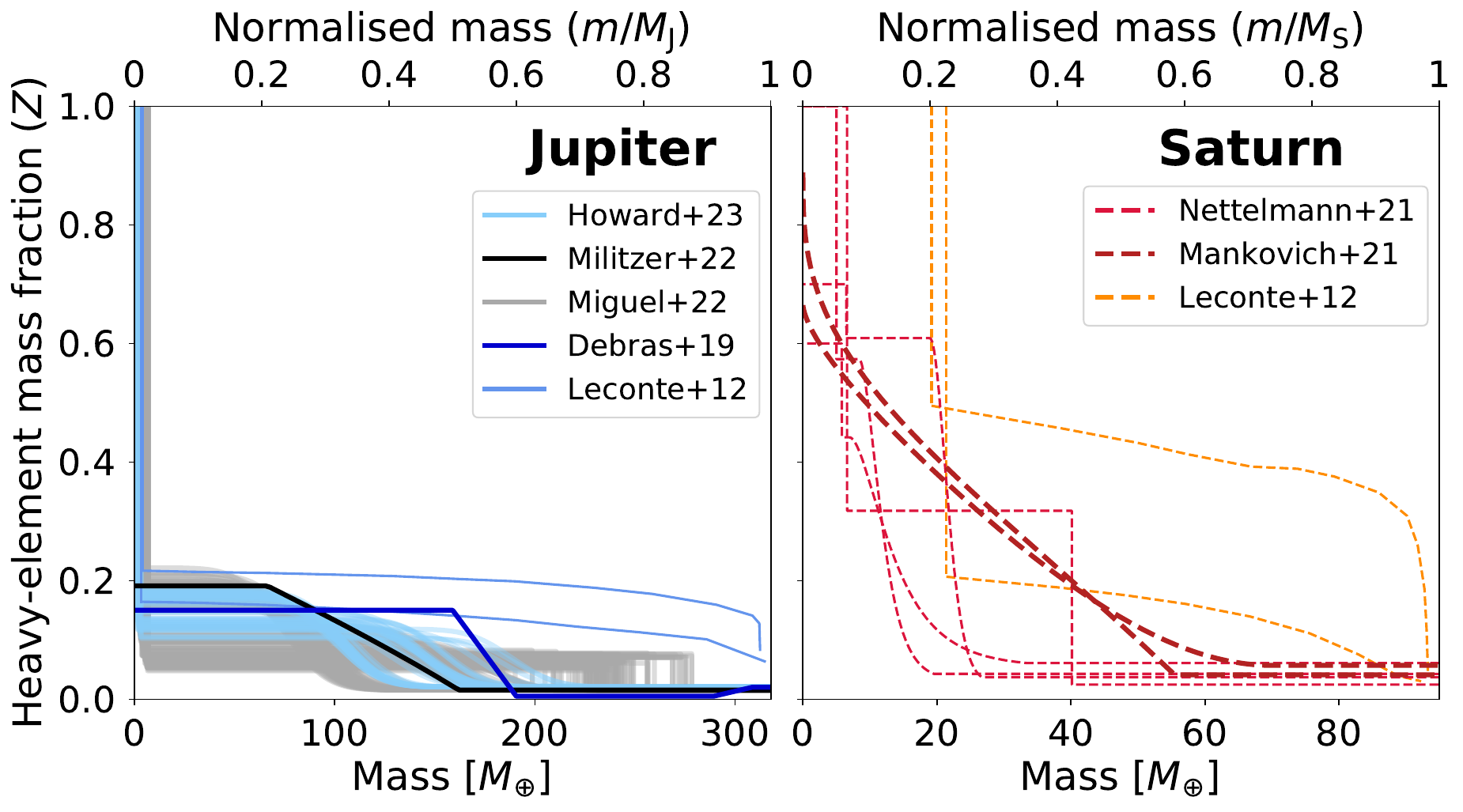}
    \caption{The heavy-element mass fraction vs.~planetary mass for Jupiter (left) and Saturn (right) from various published models. Leconte+12+Sat: \cite{Leconte2012}, Mankovich+21: \cite{2021NatAs...5.1103M}, Nettelmann+2021: \cite{nettelmann2021}, Debras+2019: \cite{Debras2019}, Howard+2023: \cite{howard2023_interior}, Miguel+2022: \cite{miguel2022}, Militzer+2022: \cite{militzer2022}.}
    \label{fig:jupsat}
  \end{center}
\end{figure}

\section{Possible Explanations for Fuzzy Cores}
There are at least four possible reasons for the failure of the old idea of a discrete, pure heavy element core surrounded by a homogeneous H-He envelope that is only mildly enriched in heavies \cite{Helled2022}.\\
The first (and oldest) explanation assumes the standard picture of delivery of the constituents but seems to fail: During formation of these planets, incoming solids (pebbles, planeteismals) evaporate and mix with the co-accreting gas, once a core exceeding of order two M$_{\oplus}$ is formed \cite{Lozovsky2017,Valletta2019}. The problem for this explanation, at least in the case of Jupiter, is that most of the mass of the planet is added by ``runaway" gas accretion (nebular inflow), a short period during which relatively little heavies are added \cite{Shibata2019,2022ApJ...926L..37S}. As a consequence, the dilution of evaporated heavies is confined to an inner region of about 20 M$_{\oplus}$ \cite{Mueller2020c}. This picture might still be relevant to Saturn, if part of its inferred fuzzy core is dominated by enhanced helium rain.

The second possible explanation appeals to mixing from traumatic events, presumably one or more giant impacts \cite{Liu2019}. The merging of massive embryos and a partially formed giant planet is also an old idea, compatible with simulations but is typically not a standard part of formation simulations 
and has been invoked to explain the observed obliquity of Jupiter. However, the likelihood of a nearly head-on collision is rather low, and most collisions are expected to be oblique. In addition, since both Jupiter and Saturn are expected to have fuzzy cores this explanation is somewhat less appealing. In any case, further investigations of this topic including high-resolution simulations are required to assess the likelihood of this scenario. 

The third possible explanation appeals to the upward stirring of heavy elements by thermal convection after planet formation. This is energetically possible 
but is very sensitive to physical parameters that are unconstrained  under Jovian conditions such as the mixing parameter, the number of layers in the case of layered convection, the opacity, and the diffusivities. 
While it is yet to be demonstrated that convective mixing in the deep interior is efficient under the post-formation conditions of Jupiter and Saturn, 
several recent studies indicate that upward mixing of heavy-elements from the deep interior is possible. Further investigations of the topic are desirable. 

The fourth explanation is linked to the formation  history.  Updated giant planet formation models show that composition gradients are a  natural outcome of the formation process \cite{Lozovsky2017,Helled2017}. In addition, recent models  suggest that significant gas accretion is initiated only when a protoplanet reaches a mass of $\sim$100 M$_{\oplus}$ and that this is a result  of an intermediate stage of efficient heavy-element accretion after core formation which provides sufficient energy to hinder rapid accretion of H-He  \cite{2018NatAs...2..873A,Venturini2020}. 
 If rapid gas accretion is indeed delayed and occurs at much higher masses, this implies that Saturn is a "failed giant planet" and has never reached runaway gas accretion. This formation path also naturally explains the fuzzy cores of Jupiter and Saturn, the differences between the bulk metallicities of Jupiter and Saturn and the mass-radius relations of observed exoplanets \cite{Helled2023}. Further investigations of this scenario are clearly desirable. 

\section{Remaining challenges in interior models}
Below, we highlight a few points that need to be addressed in order to improve our understanding of Jupiter and Saturn. \\

\noindent {\bf Unknown Temperature Profile:}\\
If we abandon, as we must, the simple assumption of an adiabatic, fully-convective interior, isentropic layers may no longer be relevant since compositional gradients may be present and these have a huge (!) effect on convective stability. 
Since the deep interior of gas giant planets  is degenerate (i.e., the temperature is small compared to the Fermi temperature of the electrons, which is of order hundreds of thousands of degrees), the thermal effect on interior models is not dominant but neither is it negligible. Doubling the internal temperature (e.g., 32,000K near the center of Jupiter instead of the old estimate of order 16,000K) trades off with an almost doubling of the heavy element mass in Jupiter (because that is small compared to the total mass). This further complicates interpretation of structure. \\

\noindent {\bf Equations of state (EoS):}\\
The inferred  interiors of Jupiter and Saturn depend on the  EoS used as small differences in the EoS, especially when it comes to hydrogen, can lead to relatively large differences in the inferred composition. As a result, it is clear that further investigations both theoretically and experimentally of materials at planetary conditions, in particular of hydrogen, and hydrogen-helium mixtures, are required. \\

\noindent {\bf The distribution of Helium:}\\
While the behavior of helium at high pressures and temperatures (i.e., the EoS) is less critical than  that of hydrogen, the distribution of He within the planet is important. He is assumed to be $\sim$ 27 $\%$ of the total H-He mass in Jupiter and Saturn (a value that is set by solar models but is consistent with cosmological estimates based on the Big Bang) and is therefore much larger than the total heavy-element mass, at least in Jupiter and possibly also in Saturn. This means that a modest redistribution of helium, likely because of limited solubility in metallic hydrogen \cite{Stevenson1975} can lead to ambiguity in the disposition of the heavier elements. In Jupiter, this effect is probably small based on the very modest atmospheric depletion of helium observed by the Galileo probe, though it does lead to a non-negligible effect on interior models and has been invoked by some to explain some layering. In Saturn, the situation is less clear since it is suspected that the atmospheric He depletion may be much larger.\\

\noindent {\bf Differentiation:}\\
Unlike the terrestrial planets, where the separation of silicates from metallic iron can be understood thermodynamically in terms of immiscibility, the separation of heavy elements from H-He in the giant planets is not thermodynamically mandated. Nor is it required by gravity, since diffusion times are longer than the age of the universe. Water and rock may mix in all proportions with metallic H at relevant temperatures.  This does not mean that layering is impossible but that ``fuzzy" cores may be present,  where the composition gradients are a result of the formation process and the possible convective mixing during the planetary evolution \cite{2018haex.bookE..44H}. 
Indeed recent formation models suggest that a  separation of elements may be possible during the early stages of planetesimal accretion where the heavy elements reach the planetary center where rocks penetrate deeper compared to water \cite{Lozovsky2017,Stevenson2022}. However, if convection is vigorous during the planetary growth, this separation could be smoothed out.  More detailed investigations of this topic are required. \\

\noindent {\bf The chemical composition of heavy elements:}\\
A remaining ambiguity concerns the exact nature of the heavy elements: In the absence of seismology, it is usually not possible to discern water from rock in interior models since our main method of observation is gravity and that does not care about composition, it only cares about the mass. Note, however, that this is only approximately true. A core of pure water will have a different effect on the observable gravity from a rock core of the same mass since they have different sizes. From hydrostatic equilibrium, this changes the pressure and density of the overlying hydrogen and helium. It is therefore still unknown what is the chemical composition of heavy elements in giant planet interiors. It is desirable to identify the sensitivity of interior model results to different assumed composition and think of new ways to constrain the heavy-element composition (e.g., moons, meteorites).

\section{The connection to exoplanets}
Jupiter and Saturn are the most explored gas giant planets and our knowledge of these two planets is critical for understanding gas giant planet in a more general sense. 
 Recent findings that Jupiter and Saturn have complex internal structure which includes inhomogeneities in composition and fuzzy cores directly affect exoplanetary science. 
This means that the over-simplified core+envelop{\bf e} or homogeneously mixed interiors must be updated and be considered when characterizing giant exoplanets.   
This in turn also affects planet evolution models that  follow the cooling and contraction of giant planets \cite{Muller2020a,Muller2020b}. While the initial conditions of adiabatic and homogeneous of giant planets are easily forgotten, this is not the case for more complex interior since they are not fully convective. Therefore, updated evolution models of giant planets should be considered. 
The consideration of non-adiabatic interiors impacts the estimates of the planetary bulk metallicity as well as the interpretation of atmospheric measurements since it is unknown how the atmospheric composition is linked to the bulk composition. In addition, we now realize that the atmospheric metallicity could even be higher than that of the deeper envelope due to "atmospheric pollution" (M{\"u}ller \& Helled, 2024), and this effect could also be important for giant exoplanets.  
Overall, determining the interiors of giant planets is important for constraining giant planet formation models, for exoplanetary characterization, and for putting  our Solar System in perspective. 
At the same time, the large number of observed giant exoplanets allow us to explore these objects in a statistical way and identify important correlations. 

\section{Summary}

Overall, despite great progress in our understanding of Jupiter and Saturn many key open questions remain, and as discussed above, several challenges in modeling the internal structure of the planets persist. \\
We suggest that future work should focus on identifying  the differences between Jupiter and Saturn and their origin. Future work should properly combine the three aspect of giant planet formation, evolution, and internal structure, while accounting for more physical processes and detailed physics. \\
In addition, the connection between the atmospheric composition of giant planets and their bulk composition  at present day must be better understood. Note that this is different from investigating atmosphere enrichment during the planetary formation process. 
While the Galileo probe measurement indicates that Jupiter's atmosphere is enriched by a factor  of three compared to solar, interior models favour more moderate enrichment in the molecular envelope. Efforts to reconcile the current-state  enriched atmosphere of Jupiter today with the predicted low-matellicity envelope are still ongoing (e.g., \cite{2023A&A...680L...2H}, M{\"u}ller \& Helled, 2024). \\
We also suggest that measuring the atmospheric composition of Saturn by an entry probe will provide an opportunity to compare the two planets. Such a probe can determine the He abundance in Saturn’s atmosphere, constraining the magnitude of He rain, and determining Saturn’s atmospheric metallicity. \\
Additional and complementary information on the internal structure of the gas giants could come from Jovian seismology. This includes the detection and characterization of  f-modes and g-modes as been detected in Saturn’s rings,  and potentially from doppler imaging.  
 Also magnetic field measurements are of great importance for constraining the interior structure. These can be used to constrain the planetary composition (via the requirement for an electrically conducting material), to determine the presence of a stably stratified layer and/or composition gradients, and for estimating the internal ohmic dissipation. Overall, magnetic field data provide key and complementary constraints for structure and dynamical models (e.g., \cite{2021MNRAS.501.2352G,2022JGRE..12707479M}).\\ 
Finally, the increasing number of gas giant exoplanets that are characterized in detail helps us to understand the nature of this planetary type and to constrain giant planet formation and evolution theory.

\section*{Conflict of Interest Statement}
The authors declare no conflicts of interest relevant to this study. 

\section*{Data Availability Statement}  
This work uses no new samples or data. 

\newpage
\subsection*{acknowledgments}
We thank Saburo Howard for valuable discussions and  technical support and the editor and reviewer for a careful reading of the paper and their valuable comments. RH acknowledges support from the Swiss National Science Foundation (SNSF) under grant \texttt{\detokenize{200020_215634}}.

\bibliographystyle{apj}

\begin{thebibliography}{0}
\expandafter\ifx\csname natexlab\endcsname\relax\def\natexlab#1{#1}\fi

\end{thebibliography}


\begin{thebibliography}{51}
\expandafter\ifx\csname natexlab\endcsname\relax\def\natexlab#1{#1}\fi

\bibitem[{{Achterberg} \& {Flasar}(2020)}]{2020PSJ.....1...30A}
{Achterberg}, R.~K. \& {Flasar}, F.~M. 2020, PSJ, 1, 30

\bibitem[{{Alibert} {et~al.}(2018){Alibert}, {Venturini}, {Helled}, {Ataiee},
  {Burn}, {Senecal}, {Benz}, {Mayer}, {Mordasini}, {Quanz}, \&
  {Sch{\"o}nb{\"a}chler}}]{2018NatAs...2..873A}
{Alibert}, Y., {Venturini}, J., {Helled}, R., {Ataiee}, S., {Burn}, R.,
  {Senecal}, L., {Benz}, W., {Mayer}, L., {Mordasini}, C., {Quanz}, S.~P., \&
  {Sch{\"o}nb{\"a}chler}, M. 2018, Nature Astronomy, 2, 873

\bibitem[{{Bolton} {et~al.}(2017{\natexlab{a}}){Bolton}, {Adriani},
  {Adumitroaie}, {Allison}, {Anderson}, {Atreya}, {Bloxham}, {Brown},
  {Connerney}, {DeJong}, {Folkner}, {Gautier}, {Grassi}, {Gulkis}, {Guillot},
  {Hansen}, {Hubbard}, {Iess}, {Ingersoll}, {Janssen}, {Jorgensen}, {Kaspi},
  {Levin}, {Li}, {Lunine}, {Miguel}, {Mura}, {Orton}, {Owen}, {Ravine},
  {Smith}, {Steffes}, {Stone}, {Stevenson}, {Thorne}, {Waite}, {Durante},
  {Ebert}, {Greathouse}, {Hue}, {Parisi}, {Szalay}, \&
  {Wilson}}]{2017Sci...356..821B}
{Bolton}, S.~J., {Adriani}, A., {Adumitroaie}, V., {Allison}, M., {Anderson},
  J., {Atreya}, S., {Bloxham}, J., {Brown}, S., {Connerney}, J.~E.~P.,
  {DeJong}, E., {Folkner}, W., {Gautier}, D., {Grassi}, D., {Gulkis}, S.,
  {Guillot}, T., {Hansen}, C., {Hubbard}, W.~B., {Iess}, L., {Ingersoll}, A.,
  {Janssen}, M., {Jorgensen}, J., {Kaspi}, Y., {Levin}, S.~M., {Li}, C.,
  {Lunine}, J., {Miguel}, Y., {Mura}, A., {Orton}, G., {Owen}, T., {Ravine},
  M., {Smith}, E., {Steffes}, P., {Stone}, E., {Stevenson}, D., {Thorne}, R.,
  {Waite}, J., {Durante}, D., {Ebert}, R.~W., {Greathouse}, T.~K., {Hue}, V.,
  {Parisi}, M., {Szalay}, J.~R., \& {Wilson}, R. 2017{\natexlab{a}}, Science,
  356, 821

\bibitem[{{Bolton} {et~al.}(2017{\natexlab{b}}){Bolton}, {Lunine}, {Stevenson},
  {Connerney}, {Levin}, {Owen}, {Bagenal}, {Gautier}, {Ingersoll}, {Orton},
  {Guillot}, {Hubbard}, {Bloxham}, {Coradini}, {Stephens}, {Mokashi}, {Thorne},
  \& {Thorpe}}]{2017SSRv..213....5B}
{Bolton}, S.~J., {Lunine}, J., {Stevenson}, D., {Connerney}, J.~E.~P., {Levin},
  S., {Owen}, T.~C., {Bagenal}, F., {Gautier}, D., {Ingersoll}, A.~P., {Orton},
  G.~S., {Guillot}, T., {Hubbard}, W., {Bloxham}, J., {Coradini}, A.,
  {Stephens}, S.~K., {Mokashi}, P., {Thorne}, R., \& {Thorpe}, R.
  2017{\natexlab{b}}, SSR, 213, 5

\bibitem[{{Chabrier} {et~al.}(2019){Chabrier}, {Mazevet}, \&
  {Soubiran}}]{Chabrier2019}
{Chabrier}, G., {Mazevet}, S., \& {Soubiran}, F. 2019, SSR, 872, 51

\bibitem[{{Debras} \& {Chabrier}(2019)}]{Debras2019}
{Debras}, F. \& {Chabrier}, G. 2019, ApJ, 872, 100

\bibitem[{{Fortney} {et~al.}(2023){Fortney}, {Militzer}, {Mankovich}, {Helled},
  {Wahl}, {Nettelmann}, {Hubbard}, {Stevenson}, {Iess}, {Marley}, \&
  {Movshovitz}}]{2023arXiv230409215F}
{Fortney}, J.~J., {Militzer}, B., {Mankovich}, C.~R., {Helled}, R., {Wahl},
  S.~M., {Nettelmann}, N., {Hubbard}, W.~B., {Stevenson}, D.~J., {Iess}, L.,
  {Marley}, M.~S., \& {Movshovitz}, N. 2023, arXiv e-prints, arXiv:2304.09215

\bibitem[{{Galanti} \& {Kaspi}(2021)}]{2021MNRAS.501.2352G}
{Galanti}, E. \& {Kaspi}, Y. 2021, MNRAS, 501, 2352

\bibitem[{{Guillot} {et~al.}(2023){Guillot}, {Fletcher}, {Helled}, {Ikoma},
  {Line}, \& {Paramentier}}]{2023ASPC..534..947G}
{Guillot}, T., {Fletcher}, L.~N., {Helled}, R., {Ikoma}, M., {Line}, M.~R., \&
  {Paramentier}, V. 2023, in Astronomical Society of the Pacific Conference
  Series, Vol. 534, Astronomical Society of the Pacific Conference Series, ed.
  S.~{Inutsuka}, Y.~{Aikawa}, T.~{Muto}, K.~{Tomida}, \& M.~{Tamura}, 947

\bibitem[{{Guillot} {et~al.}(2004){Guillot}, {Stevenson}, {Hubbard}, \&
  {Saumon}}]{Guillot2004}
{Guillot}, T., {Stevenson}, D.~J., {Hubbard}, W.~B., \& {Saumon}, D. {The
  interior of Jupiter}, Vol.~1 (Cambridge University Press), 35--57

\bibitem[{{Helled}(2023)}]{Helled2023}
{Helled}, R. 2023, AAP, 675, L8

\bibitem[{{Helled} {et~al.}(2014){Helled}, {Bodenheimer}, {Podolak}, {Boley},
  {Meru}, {Nayakshin}, {Fortney}, {Mayer}, {Alibert}, \&
  {Boss}}]{2014prpl.conf..643H}
{Helled}, R., {Bodenheimer}, P., {Podolak}, M., {Boley}, A., {Meru}, F.,
  {Nayakshin}, S., {Fortney}, J.~J., {Mayer}, L., {Alibert}, Y., \& {Boss},
  A.~P. 2014, in Protostars and Planets VI, ed. H.~{Beuther}, R.~S. {Klessen},
  C.~P. {Dullemond}, \& T.~{Henning}, 643--665

\bibitem[{{Helled} \& {Guillot}(2018)}]{2018haex.bookE..44H}
{Helled}, R. \& {Guillot}, T. 2018, in Handbook of Exoplanets, ed. H.~J. {Deeg}
  \& J.~A. {Belmonte}, 44

\bibitem[{{Helled} {et~al.}(2020){Helled}, {Mazzola}, \&
  {Redmer}}]{2020NatRP...2..562H}
{Helled}, R., {Mazzola}, G., \& {Redmer}, R. 2020, Nature Reviews Physics, 2,
  562

\bibitem[{{Helled} \& {Morbidelli}(2021)}]{2021exbi.book...12H}
{Helled}, R. \& {Morbidelli}, A. 2021, in ExoFrontiers; Big Questions in
  Exoplanetary Science, ed. N.~{Madhusudhan}, 12--1

\bibitem[{{Helled} {et~al.}(2022{\natexlab{a}}){Helled}, {Movshovitz}, \&
  {Nettelmann}}]{Helled2022b}
{Helled}, R., {Movshovitz}, N., \& {Nettelmann}, N. 2022{\natexlab{a}}, arXiv
  e-prints, arXiv:2202.10046

\bibitem[{{Helled} {et~al.}(2022{\natexlab{b}}){Helled}, {Movshovitz}, \&
  {Nettelmann}}]{2022arXiv220210046H}
---. 2022{\natexlab{b}}, arXiv e-prints, arXiv:2202.10046

\bibitem[{{Helled} \& {Stevenson}(2017)}]{Helled2017}
{Helled}, R. \& {Stevenson}, D. 2017, ApJl, 840, L4

\bibitem[{{Helled} {et~al.}(2022{\natexlab{c}}){Helled}, {Stevenson}, {Lunine},
  {Bolton}, {Nettelmann}, {Atreya}, {Guillot}, {Militzer}, {Miguel}, \&
  {Hubbard}}]{Helled2022}
{Helled}, R., {Stevenson}, D.~J., {Lunine}, J.~I., {Bolton}, S.~J.,
  {Nettelmann}, N., {Atreya}, S., {Guillot}, T., {Militzer}, B., {Miguel}, Y.,
  \& {Hubbard}, W.~B. 2022{\natexlab{c}}, Icarus, 378, 114937

\bibitem[{{Howard} {et~al.}(2023{\natexlab{a}}){Howard}, {Guillot}, {Bazot},
  {Miguel}, {Stevenson}, {Galanti}, {Kaspi}, {Hubbard}, {Militzer}, {Helled},
  {Nettelmann}, {Idini}, \& {Bolton}}]{howard2023_interior}
{Howard}, S., {Guillot}, T., {Bazot}, M., {Miguel}, Y., {Stevenson}, D.~J.,
  {Galanti}, E., {Kaspi}, Y., {Hubbard}, W.~B., {Militzer}, B., {Helled}, R.,
  {Nettelmann}, N., {Idini}, B., \& {Bolton}, S. 2023{\natexlab{a}}, AAP, 672,
  A33

\bibitem[{{Howard} {et~al.}(2023{\natexlab{b}}){Howard}, {Guillot}, {Markham},
  {Helled}, {M{\"u}ller}, {Stevenson}, {Lunine}, {Miguel}, \&
  {Nettelmann}}]{2023A&A...680L...2H}
{Howard}, S., {Guillot}, T., {Markham}, S., {Helled}, R., {M{\"u}ller}, S.,
  {Stevenson}, D.~J., {Lunine}, J.~I., {Miguel}, Y., \& {Nettelmann}, N.
  2023{\natexlab{b}}, AAP, 680, L2

\bibitem[{{Iess} {et~al.}(2019){Iess}, {Militzer}, {Kaspi}, {Nicholson},
  {Durante}, {Racioppa}, {Anabtawi}, {Galanti}, {Hubbard}, {Mariani},
  {Tortora}, {Wahl}, \& {Zannoni}}]{2019Sci...364.2965I}
{Iess}, L., {Militzer}, B., {Kaspi}, Y., {Nicholson}, P., {Durante}, D.,
  {Racioppa}, P., {Anabtawi}, A., {Galanti}, E., {Hubbard}, W., {Mariani},
  M.~J., {Tortora}, P., {Wahl}, S., \& {Zannoni}, M. 2019, Science, 364,
  aat2965

\bibitem[{{Leconte} \& {Chabrier}(2012)}]{Leconte2012}
{Leconte}, J. \& {Chabrier}, G. 2012, AAP, 540, A20

\bibitem[{{Leconte} \& {Chabrier}(2013)}]{Leconte2013}. 2013, Nature Geoscience, 6, 347

\bibitem[{{Liu} {et~al.}(2019){Liu}, {Hori}, {M{\"u}ller}, {Zheng}, {Helled},
  {Lin}, \& {Isella}}]{Liu2019}
{Liu}, S.-F., {Hori}, Y., {M{\"u}ller}, S., {Zheng}, X., {Helled}, R., {Lin},
  D., \& {Isella}, A. 2019, \nat, 572, 355

\bibitem[{{Lozovsky} {et~al.}(2017){Lozovsky}, {Helled}, {Rosenberg}, \&
  {Bodenheimer}}]{Lozovsky2017}
{Lozovsky}, M., {Helled}, R., {Rosenberg}, E.~D., \& {Bodenheimer}, P. 2017,
  ApJ, 836, 227

\bibitem[{{Mankovich}(2020)}]{Mankovich2020AGU}
{Mankovich}, C.~R. 2020, AGU Advances, 1, e00142

\bibitem[{{Mankovich} \& {Fortney}(2020)}]{Mankovich2020}
{Mankovich}, C.~R. \& {Fortney}, J.~J. 2020, ApJ, 889, 51

\bibitem[{{Mankovich} \& {Fuller}(2021)}]{2021NatAs...5.1103M}
{Mankovich}, C.~R. \& {Fuller}, J. 2021, Nature Astronomy, 5, 1103

\bibitem[{{Miguel} {et~al.}(2022){Miguel}, {Bazot}, {Guillot}, {Howard},
  {Galanti}, {Kaspi}, {Hubbard}, {Militzer}, {Helled}, {Atreya}, {Connerney},
  {Durante}, {Kulowski}, {Lunine}, {Stevenson}, \& {Bolton}}]{miguel2022}
{Miguel}, Y., {Bazot}, M., {Guillot}, T., {Howard}, S., {Galanti}, E., {Kaspi},
  Y., {Hubbard}, W.~B., {Militzer}, B., {Helled}, R., {Atreya}, S.~K.,
  {Connerney}, J.~E.~P., {Durante}, D., {Kulowski}, L., {Lunine}, J.~I.,
  {Stevenson}, D., \& {Bolton}, S. 2022, AAP, 662, A18

\bibitem[{{Miguel} \& {Vazan}(2023)}]{MiguelVazan2023}
{Miguel}, Y. \& {Vazan}, A. 2023, Remote Sensing, 15, 681

\bibitem[{{Militzer} {et~al.}(2022){Militzer}, {Hubbard}, {Wahl}, {Lunine},
  {Galanti}, {Kaspi}, {Miguel}, {Guillot}, {Moore}, {Parisi}, {Connerney},
  {Helled}, {Cao}, {Mankovich}, {Stevenson}, {Park}, {Wong}, {Atreya},
  {Anderson}, \& {Bolton}}]{militzer2022}
{Militzer}, B., {Hubbard}, W.~B., {Wahl}, S., {Lunine}, J.~I., {Galanti}, E.,
  {Kaspi}, Y., {Miguel}, Y., {Guillot}, T., {Moore}, K.~M., {Parisi}, M.,
  {Connerney}, J. E.~P., {Helled}, R., {Cao}, H., {Mankovich}, C., {Stevenson},
  D.~J., {Park}, R.~S., {Wong}, M., {Atreya}, S.~K., {Anderson}, J., \&
  {Bolton}, S.~J. 2022, PSJ, 3, 185

\bibitem[{{Militzer} {et~al.}(2019){Militzer}, {Wahl}, \&
  {Hubbard}}]{2019ApJ...879...78M}
{Militzer}, B., {Wahl}, S., \& {Hubbard}, W.~B. 2019, ApJ, 879, 78

\bibitem[{{Mizuno}(1980)}]{1980PThPh..64..544M}
{Mizuno}, H. 1980, Progress of Theoretical Physics, 64, 544

\bibitem[{{Moore} {et~al.}(2022){Moore}, {Barik}, {Stanley}, {Stevenson},
  {Nettelmann}, {Helled}, {Guillot}, {Militzer}, \&
  {Bolton}}]{2022JGRE..12707479M}
{Moore}, K.~M., {Barik}, A., {Stanley}, S., {Stevenson}, D.~J., {Nettelmann},
  N., {Helled}, R., {Guillot}, T., {Militzer}, B., \& {Bolton}, S. 2022,
  Journal of Geophysical Research (Planets), 127, e2022JE007479

\bibitem[{{Movshovitz} {et~al.}(2020){Movshovitz}, {Fortney}, {Mankovich},
  {Thorngren}, \& {Helled}}]{2020ApJ...891..109M}
{Movshovitz}, N., {Fortney}, J.~J., {Mankovich}, C., {Thorngren}, D., \&
  {Helled}, R. 2020, ApJ, 891, 109

\bibitem[{{M{\"u}ller} {et~al.}(2020{\natexlab{a}}){M{\"u}ller}, {Ben-Yami}, \&
  {Helled}}]{Muller2020a}
{M{\"u}ller}, S., {Ben-Yami}, M., \& {Helled}, R. 2020{\natexlab{a}}, ApJ,
  903, 147

\bibitem[{{M{\"u}ller} {et~al.}(2020{\natexlab{b}}){M{\"u}ller}, {Helled}, \&
  {Cumming}}]{Mueller2020c}
{M{\"u}ller}, S., {Helled}, R., \& {Cumming}, A. 2020{\natexlab{b}}, AAP, 638,
  A121

\bibitem[{{M{\"u}ller} {et~al.}(2020{\natexlab{c}}){M{\"u}ller}, {Helled}, \&
  {Cumming}}]{Muller2020b}
---. 2020{\natexlab{c}}, AAP, 638, A121

\bibitem[{{Nettelmann} {et~al.}(2021){Nettelmann}, {Movshovitz}, {Ni},
  {Fortney}, {Galanti}, {Kaspi}, {Helled}, {Mankovich}, \&
  {Bolton}}]{nettelmann2021}
{Nettelmann}, N., {Movshovitz}, N., {Ni}, D., {Fortney}, J.~J., {Galanti}, E.,
  {Kaspi}, Y., {Helled}, R., {Mankovich}, C.~R., \& {Bolton}, S. 2021, PSJ, 2,
  241

\bibitem[{{Ni}(2020)}]{2020A&A...639A..10N}
{Ni}, D. 2020, AAP, 639, A10

\bibitem[{{Pollack} {et~al.}(1996){Pollack}, {Hubickyj}, {Bodenheimer},
  {Lissauer}, {Podolak}, \& {Greenzweig}}]{1996Icar..124...62P}
{Pollack}, J.~B., {Hubickyj}, O., {Bodenheimer}, P., {Lissauer}, J.~J.,
  {Podolak}, M., \& {Greenzweig}, Y. 1996, Icarus, 124, 62

\bibitem[{{Shibata} \& {Helled}(2022)}]{2022ApJ...926L..37S}
{Shibata}, S. \& {Helled}, R. 2022, ApJl, 926, L37

\bibitem[{{Shibata} \& {Ikoma}(2019)}]{Shibata2019}
{Shibata}, S. \& {Ikoma}, M. 2019, MNRAS, 487, 4510

\bibitem[{{Stevenson}(1975)}]{Stevenson1975}
{Stevenson}, D.~J. 1975, \prb, 12, 3999

\bibitem[{{Stevenson} {et~al.}(2022{\natexlab{a}}){Stevenson}, {Bodenheimer},
  {Lissauer}, \& {D'Angelo}}]{Stevenson22}
{Stevenson}, D.~J., {Bodenheimer}, P., {Lissauer}, J.~J., \& {D'Angelo}, G.
  2022{\natexlab{a}}, PSJ, 3, 74

\bibitem[{{Stevenson} {et~al.}(2022{\natexlab{b}}){Stevenson}, {Bodenheimer},
  {Lissauer}, \& {D'Angelo}}]{Stevenson2022}
---. 2022{\natexlab{b}}, PSJ, 3, 74

\bibitem[{{Valletta} \& {Helled}(2019)}]{Valletta2019}
{Valletta}, C. \& {Helled}, R. 2019, ApJ, 871, 127

\bibitem[{{Vazan} {et~al.}(2018){Vazan}, {Helled}, \& {Guillot}}]{vazan2018}
{Vazan}, A., {Helled}, R., \& {Guillot}, T. 2018, AAP, 610, L14

\bibitem[{{Venturini} \& {Helled}(2020)}]{Venturini2020}
{Venturini}, J. \& {Helled}, R. 2020, AAP, 634, A31

\bibitem[{{Wahl} {et~al.}(2017){Wahl}, {Hubbard}, {Militzer}, {Guillot},
  {Miguel}, {Movshovitz}, {Kaspi}, {Helled}, {Reese}, {Galanti}, {Levin},
  {Connerney}, \& {Bolton}}]{wahl2017}
{Wahl}, S.~M., {Hubbard}, W.~B., {Militzer}, B., {Guillot}, T., {Miguel}, Y.,
  {Movshovitz}, N., {Kaspi}, Y., {Helled}, R., {Reese}, D., {Galanti}, E.,
  {Levin}, S., {Connerney}, J.~E., \& {Bolton}, S.~J. 2017, \grl, 44, 4649

\end{thebibliography}

\end{document}